\documentclass[runningheads]{llncs}
\usepackage[T1]{fontenc}
\usepackage{graphicx}

\usepackage{rotating}
\usepackage{booktabs}
\usepackage{makecell}
\usepackage{comment}
\usepackage{caption}

\usepackage{hyperref}
\usepackage{color}

\begin{document}
\title{Improving Virtual Contrast Enhancement using Longitudinal Data}
\titlerunning{Improving Virtual Contrast Enhancement using Longitudinal Data}
%

\author{
Pierre Fayolle\inst{1,2,5} \and
Alexandre Bône\inst{1} \and 
Noëlie Debs\inst{1} \and 
Philippe Robert\inst{1} \and 
Pascal Bourdon\inst{4,5} \and 
Rémy Guillevin\inst{2,3,5} \and 
David Helbert\inst{4,5}}
\authorrunning{P. Fayolle et al.}
%
\institute{Guerbet Research, Villepinte, France \and
LMA, Université de Poitiers, France \and
CHU de Poitiers, Poitiers, France \and
XLIM, Université de Poitiers, France \and
I3M, Common Laboratory CNRS-Siemens, France}

%
\maketitle              
\begin{abstract}
Gadolinium-based contrast agents (GBCAs) are widely used in magnetic resonance imaging (MRI) to enhance lesion detection and characterisation, particularly in the field of neuro-oncology.
Nevertheless, concerns regarding gadolinium retention and accumulation in brain and body tissues, most notably for diseases that require close monitoring and frequent GBCA injection, have led to the need for strategies to reduce dosage.

In this study, a deep learning framework is proposed for the virtual contrast enhancement of full-dose post-contrast T1-weighted MRI images from corresponding low-dose acquisitions.
The contribution of the presented model is its utilisation of longitudinal information, which is achieved by incorporating a prior full-dose MRI examination from the same patient.
A comparative evaluation against a non-longitudinal single session model demonstrated that the longitudinal approach significantly improves image quality across multiple reconstruction metrics.
Furthermore, experiments with varying simulated contrast doses confirmed the robustness of the proposed method.
These results emphasize the potential of integrating prior imaging history into deep learning-based virtual contrast enhancement pipelines to reduce GBCA usage without compromising diagnostic utility, thus paving the way for safer, more sustainable longitudinal monitoring in clinical MRI practice.

\end{abstract}
%
%
\section{Introduction}
Magnetic Resonance Imaging (MRI) is an essential diagnostic tool in clinical practice, offering non-invasive, high-resolution visualization of anatomical and functional structures.
Contrast-enhanced MRI, particularly T1-weighted imaging following the administration of gadolinium-based contrast agents (GBCAs), has been shown to significantly improve lesion detection and characterisation, especially in the fields of neuroimaging and oncology~\cite{runge2000safety}.
However, concerns have been raised regarding the long-term safety of GBCAs, including gadolinium retention and accumulation in brain and body tissues, even in patients with normal renal function~\cite{gulani2017gadolinium,kanda2014high}.
These issues are of particular significance for patients afflicted with brain tumours or other chronic conditions, who require regular follow-up examinations involving contrast-enhanced MRI.

In order to address the aforementioned risks, a critical research objective has emerged: the reduction of the dose of contrast agent without compromising image quality.
However, lower doses typically result in diminished signal enhancement and impaired diagnostic performance~\cite{gong2018deep}.
Recently, deep learning–based approaches have emerged as powerful tools for virtually enhancing medical images from low-dose acquisitions or simulations~\cite{ammari2022can,gong2018deep,haase2023reduction,luo2021deep,mallio2023artificial,pinetz2024gadolinium}.
These techniques have been demonstrated to be capable of effectively learning spatial and contrast relationships in paired data, with the potential to significantly reduce contrast agent usage.
Although synthesized contrast-enhanced images appear realistic, reader studies have revealed critical limitations, including missed subtle lesions, hallucinated findings~\cite{ammari2022can,haase2023reduction,luo2021deep}, and excessive smoothing~\cite{ammari2022can}, which prevent their use in clinical routine.

A fundamental innovation of our approach is the utilisation of not only the current low-dose MRI, but also the patient's previous MRI exam acquired at full contrast dose.
The model is trained on paired MRI volumes composed of pre- and post-contrast images from two sequential MRI sessions: one with a standard 100\% dose (Session 1) and another with an ajustable synthetic low-dose (Session 2).
The longitudinal design of the model might facilitate the acquisition of subject-specific contrast enhancement patterns and anatomical knowledge, thereby enhancing its capacity to synthesize high-quality full-dose contrast images.
The objective of this study is to evaluate whether incorporating prior clinical imaging data through a deep learning–based virtual contrast enhancement approach leads to improved image fidelity compared to conventional non-longitudinal methods, thereby supporting the feasibility of reducing GBCA usage.

\section{Materials and Methods}
\begin{figure}
    \centering
    \includegraphics[width=\textwidth]{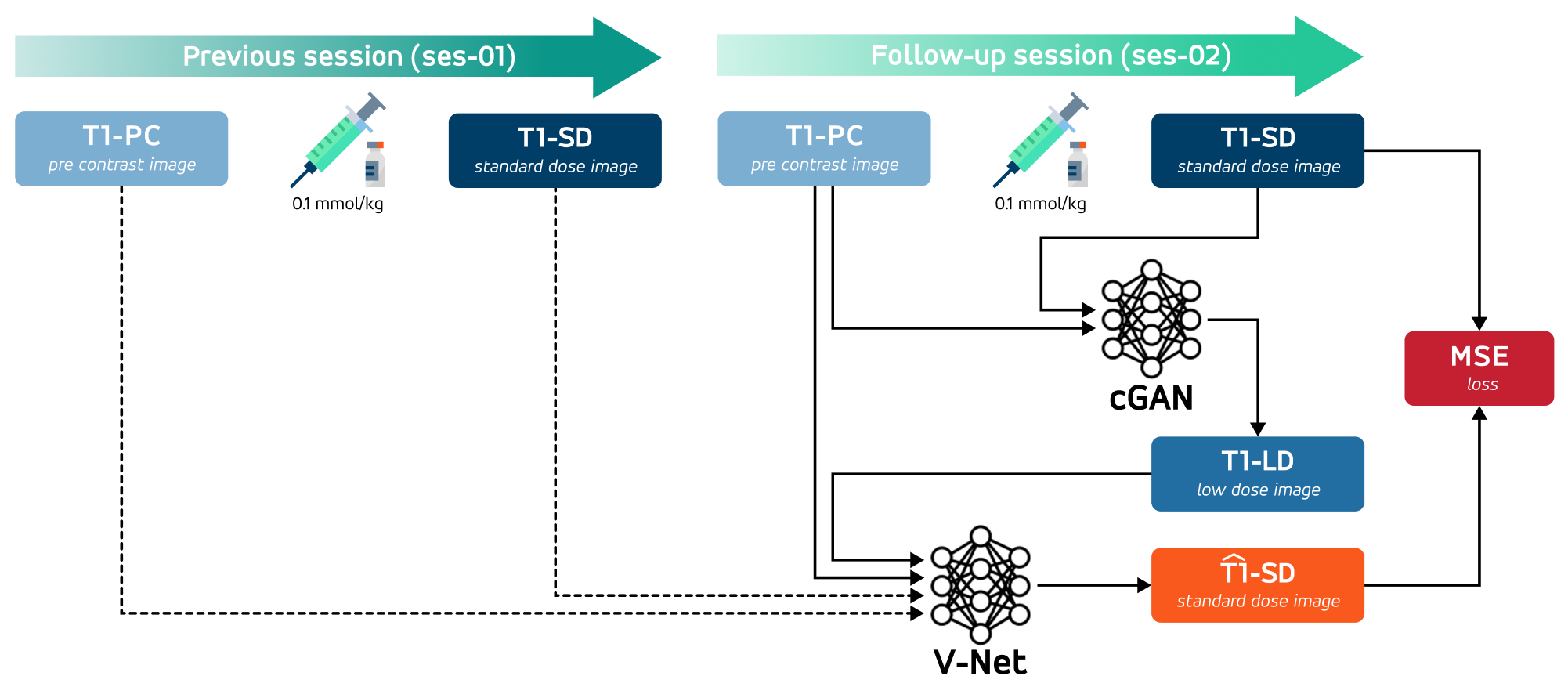}
    \caption{Workflow of the Proposed Longitudinal Virtual Contrast Enhancement Model for T1-Weighted Images. Dotted arrow shows the main contribution of the proposed work. cGAN: conditional Generative Adversarial Networks, MSE: Mean Square Error.}
    \label{fig_workflow}
\end{figure}
\subsection{Longitudinal Virtual Contrast Enhancement Workflow}
Figure~\ref{fig_workflow}  shows an overview of the proposed longitudinal virtual contrast enhancement method for standard-dose post-contrast T1-weighted (T1-SD) MRI images.
This method uses prior session data and low-dose acquisitions from a follow-up session.
The diagram is divided into two temporal phases: the previous session (ses-01) and the follow-up session (ses-02). 
In the previous session, a pre-contrast T1-weighted image (T1-PC) is acquired prior to the administration of 0.1 mmol/kg gadolinium contrast agent, followed by the acquisition of T1-SD.
These images thus serve as high-quality historical references.
In the follow-up session (ses-02),  a new T1-PC image is acquired and a standard dose is administered in a manner similar to that of ses-01.
However, in the proposed framework, a simulation of a low-dose equivalent (T1-LD) from the follow-up T1-PC and T1-SD is conducted using a conditional generative adversarial network (cGAN). 
This synthetic low-dose images are generated by simulating contrast levels corresponding to 10\%, 15\%, 20\%, 25\% and 33\% of the standard gadolinium dose.
The T1-PC and T1-LD from ses-02, along with the T1-PC and T1-SD from ses-01, are then subsequently concatenated and provided as multi-channel inputs to a 3D V-Net architecture.
This network is trained to generate an image that approximates the full-dose post-contrast image $\widehat{\mathrm{T1}}$-SD for the follow-up session.
The predicted image is then compared to the actual follow-up T1-SD using a mean squared error (MSE) loss, enforcing voxel-wise intensity fidelity.

\subsection{Public dataset}
This study uses images from the ACRIN-DSC-MR-Brain collection~\cite{kinahan2019data}, a publicly available dataset hosted on The Cancer Imaging Archive (TCIA) at the following link: \url{https://www.cancerimagingarchive.net/collection/acrin-dsc-mr-brain/}.\\
This longitudinal dataset includes 123 patients diagnosed with recurrent glioblastoma undergoing anti-angiogenic therapy. For this work, only the axial pre- and post-contrast T1-weighted images from the two most recent imaging sessions available for each subject were used.
A total of 26 patients were excluded from the dataset due to the absence of suitable pre- and post-contrast T1-weighted images from two separate sessions.
Of the remaining 97 patients left, the average time between two sessions is 57.21\,$\pm$\,31.72 days, ranging from a minimum of 16 days to a maximum of 312 days.
The dataset was split into 75 subjects for training, 5 for validation, and 17 for testing.

\subsection{Data Preprocessing}
The two sessions for each subject were processed sequentially. For both T1-PC and T1-SD images, skull stripping was performed using HD-BET~\cite{isensee2019automated}.
Prior to that, all volumes were cropped to the brain region and resampled to isotropic 1 mm$^{3}$ resolution.
The T1-SD image from the ses-01 was elastically registered to the corresponding T1-PC image using SimpleITK extension SimpleElastix~\cite{marstal2016simpleelastix}, and both were min-max normalized jointly to preserve relative intensity scales.
For the ses-02, T1-PC and T1-SD images were first aligned to the T1-PC from the ses-01 using the same registration method to ensure inter-session correspondence.
The same crop box from the ses-01 was applied to the ses-02 volumes for consistent spatial coverage.
As the ACRIN dataset does not provide T1-LD, a synthetic contrast reduction approach was used, as proposed in~\cite{pinetz2023faithful}, to simulate adjustable dose T1ce images from the ses-02.
Subsequently, all images from the ses-02 (T1-PC, T1-SD, and T1-LD) were jointly normalized using the identical strategy employed in the preceding session.
Following these preprocessing steps, the four resulting volumes, T1-PC and T1-SD from the ses-01, with T1-PC and T1-LD from the ses-02, were stacked along the channel dimension to form an input of size 160x192x160x4.

To improve the model generalization and to keep the input dimensions consistent, a complete data augmentation pipeline was used during training.
Spatial variability was introduced through random 3D flips along each anatomical axis (applied with a probability of 0.5 per axis) and random affine transformations, including small rotations (up to 0.05 radians), translations (up to 5 voxels), and uniform scaling within a \,$\pm$\,10\% range.
In addition to spatial augmentations, intensity-based transformations were applied to increase robustness to signal variations.
These included the addition of random Gaussian noise (standard deviation of 0.01, applied with 30\% probability) and random intensity shifts (offset of 0.1, applied with 50\% probability).

\subsection{Training Strategy}
We trained a 3D V-Net model proposed in \cite{bone2021contrast} and originally adapted from \cite{milletari2016v} on whole-brain volumes.
The network was trained for 500 epochs with a batch size of 1 and an initial learning rate of $10^{-4}$.
Optimization was performed using the Adam optimizer. To adapt the learning rate during training, we employed a scheduler that reduced the learning rate by a factor of 0.5 when the loss stabilized for 10 consecutive epochs.
This dynamic adjustment encouraged more stable convergence throughout training.

\subsection{Model Evaluation}
To assess the effectiveness of the proposed longitudinal model, its performance was compared to that of a single session model that was trained only on data from the ses-02, called the single session model.
This approach aligns with the strategies commonly adopted in recent studies aimed at reducing contrast doses~\cite{gong2018deep,haase2023artificial,mingo2023amplifying,pasumarthi2021generic,pinetz2024gadolinium}. 
Importantly, both the models were trained using the same architecture, loss functions, and hyperparameters, ensuring a fair comparison.
For that purpose, three different reconstruction metrics were used. The Mean Square Error (MSE, $\times 10^{-2}$), the Peak Signal-to-Noise Ratio (PSNR, dB) and the SSIM (Structural SIMilarity).
Appropriate paired tests were conducted for each metric to assess the statistical significance of the difference between the two models.
This comparative evaluation provides a clearer insight into the impact of longitudinal information on the virtual contrast enhancement quality.
Each training was performed on a NVIDIA Tesla T4 with 16Gb of RAM, taking approximatively 23 hours for the single session model and 25 hours for the longitudinal model.
Code will be shared upon acceptance of our work.

\section{Results}

\begin{table}[t!]
\centering
\renewcommand{\arraystretch}{1.3} 
\setlength{\tabcolsep}{12pt}
\caption{Quantitative Results Between Low-Dose T1 (T1-LD) Images, Single Session and Longitudinal Models Across Reconstruction Metrics Compared with Full-Dose T1 (ses-02 T1-SD) Ground-Truth.}\label{tab_metrics}
\begin{tabular}{l|c|c|c}
\toprule
\textbf{Model} & \makecell{\textbf{MSE} ($\times 10^{-2}$) ($\downarrow$)} & \makecell{\textbf{PSNR} (dB) ($\uparrow$)} & \makecell{\textbf{SSIM} ($\uparrow$)} \\
\midrule
\midrule
T1-LD & 0.2211\,$\pm$\,0.2292 & 28.2346\,$\pm$\,3.7540 & 0.9315\,$\pm$\,0.0223\\
Single Session & 0.1564\,$\pm$\,0.1190 & 29.1920\,$\pm$\,3.4279 & 0.9317\,$\pm$\,0.0166\\
Longitudinal & \textbf{0.1160}\,$\pm$\,0.1555 & \textbf{32.6337}\,$\pm$\,5.9517 & \textbf{0.9726}\,$\pm$\,0.0177\\
      & $p =$ 0.0569* & $p =$ 0.0083* & $p <$ 0.0001* \\
\bottomrule
\end{tabular}
\captionsetup{justification=justified}
\caption*{\footnotesize *The reported $p$-values indicate the statistical significance of improvements in the longitudinal model compared to the single session baseline.}
\end{table}

\subsection{Comparative Quantitative Analysis}
Table~\ref{tab_metrics} shows the quantitative results of the single session model compared to the longitudinal model at 25\% dose.
For the single session model, the MSE was 0.1564\,$\pm$\,0.1190, PSNR was 29.1920\,$\pm$\,3.4279, and the SSIM was 0.9317\,$\pm$\,0.0166.
In contrast, the longitudinal model yielded improved performance with an MSE of 0.1160\,$\pm$\,0.1555, PSNR of 32.6337\,$\pm$\,5.9517, and SSIM of 0.9726\,$\pm$\,0.0177.

For MSE, a Shapiro-Wilk test indicated non-normality ($p = 0.0018$), and the Wilcoxon signed-rank test was therefore used, revealing a trend toward significance ($p = 0.0569$).
For PSNR and SSIM, normality was confirmed ($p = 0.9645$ and $p = 0.3069$, respectively), and paired t-tests were applied. The PSNR comparison yielded a statistically significant improvement ($t = -3.01$, $p = 0.0083$), and SSIM showed a highly significant difference ($t = -11.73$, $p = 2.85\times10^{-9}$).
These results support the conclusion that incorporating longitudinal information substantially improves image quality across multiple quantitative metrics.

\begin{figure}
    \centering
    \includegraphics[width=\textwidth]{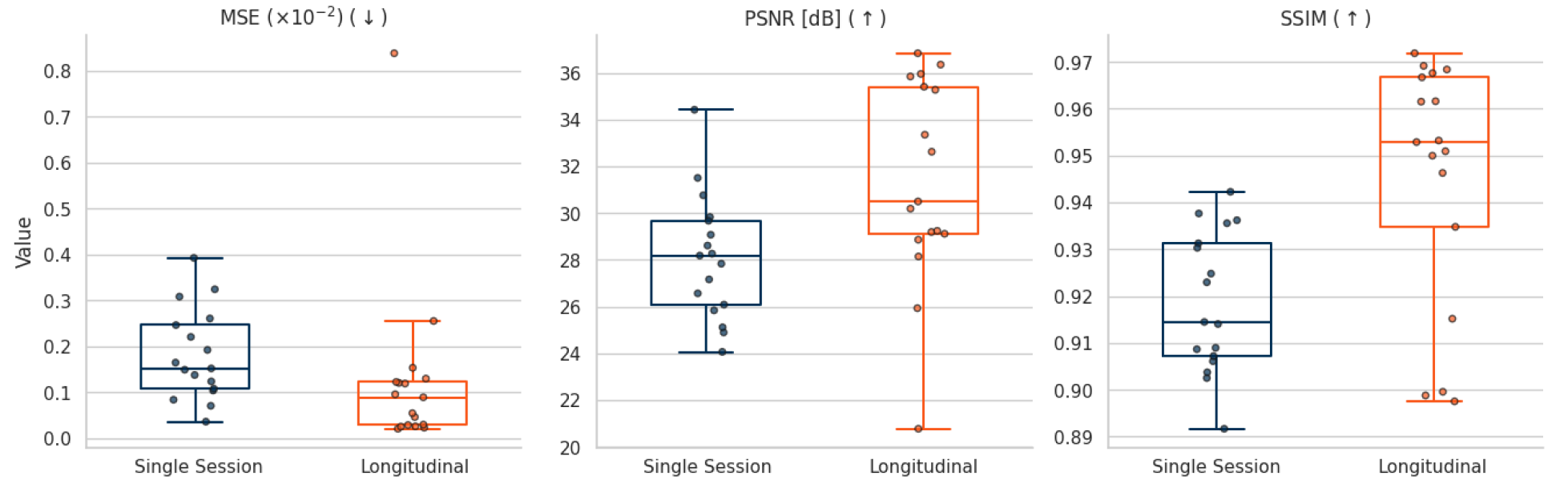}
    \caption{Boxplot Analysis of Metrics Between the Single Session and Longitudinal Models. MSE: Mean Square Error, PSNR: Peak Signal-to-Noise Ratio, SSIM: Structural Similarity Index.}
    \label{fig_boxplots}
\end{figure}

To visually illustrate these differences, Figure~\ref{fig_boxplots} presents boxplots comparing the two models across the three metrics.
The distributions clearly highlight the reduced variance and overall improvement in performance when longitudinal information is integrated.

\begin{figure}
    \centering
    \includegraphics[width=0.9\textwidth]{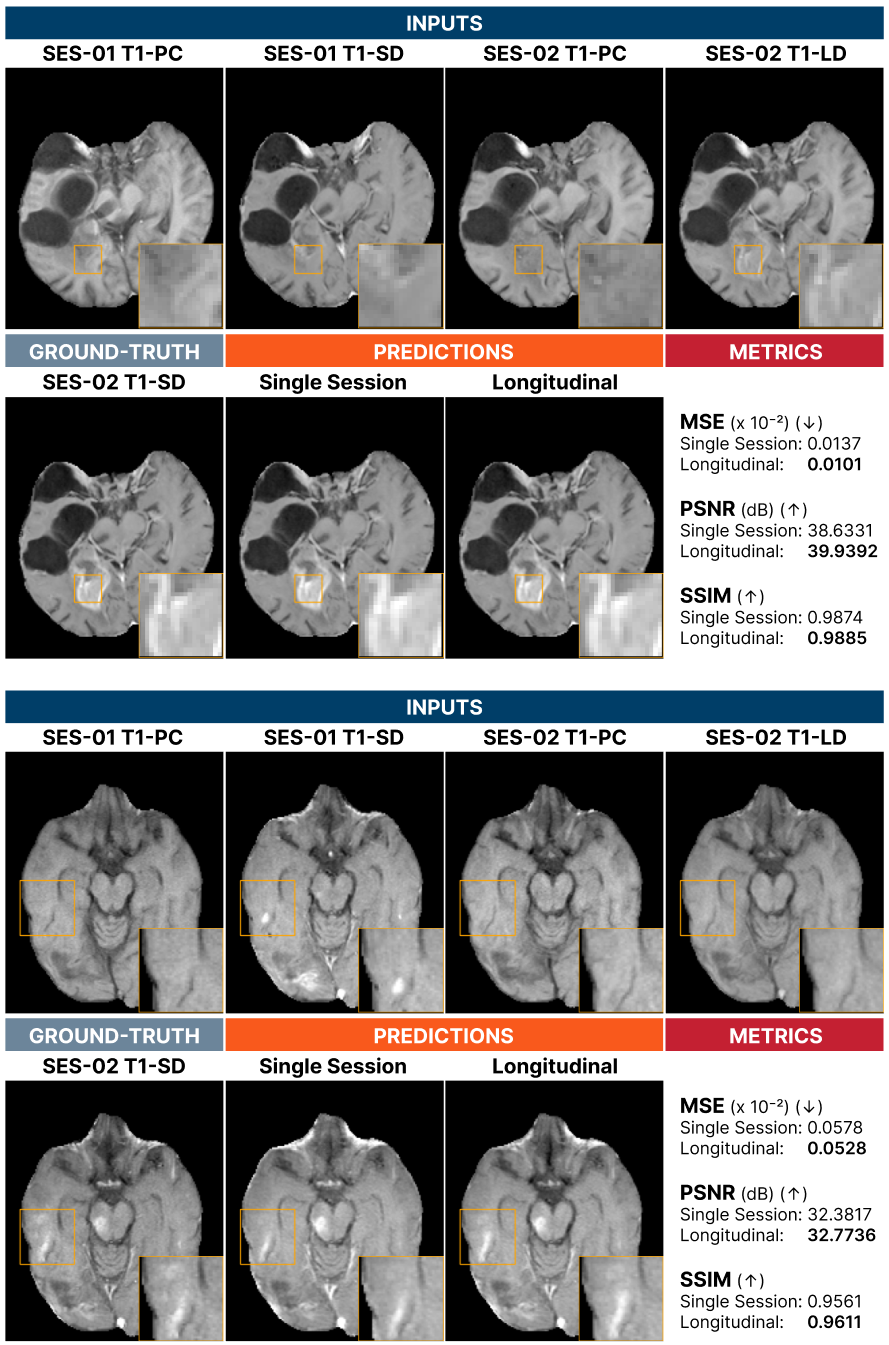}
    \caption{Comparison Between Single Session and Longitudinal Data-Driven Models. PC: Pre-contrast, SD: Standard-dose, LD: Low-dose (25\% synthetized dose).}
    \label{fig_quali}
\end{figure} 

\subsection{Qualitative comparison}
As illustrated in Figure~\ref{fig_quali}, representative lesion-centered slices from two distinct test subjects are presented for a 25\% synthetized dose, thereby enabling a qualitative comparison between the single session and the proposed longitudinal models.
In all cases, both methods were found to be equally successful in virtually enhancing contrast in anatomically coherent images, with consistent lesion localisation and structural delineation.
It is noteworthy that visual inspection reveals only subtle differences between the two approaches, particularly in regions of lesion enhancement and surrounding tissue contrast.
These observations are consistent with the robust performance of both models and corroborate the quantitative findings, wherein the longitudinal model demonstrates statistically significant improvements despite minimal perceptible differences in visual appearance.

\subsection{Dose Variation Evaluation}

\begin{figure}[t!]
    \centering
    \includegraphics[width=\textwidth]{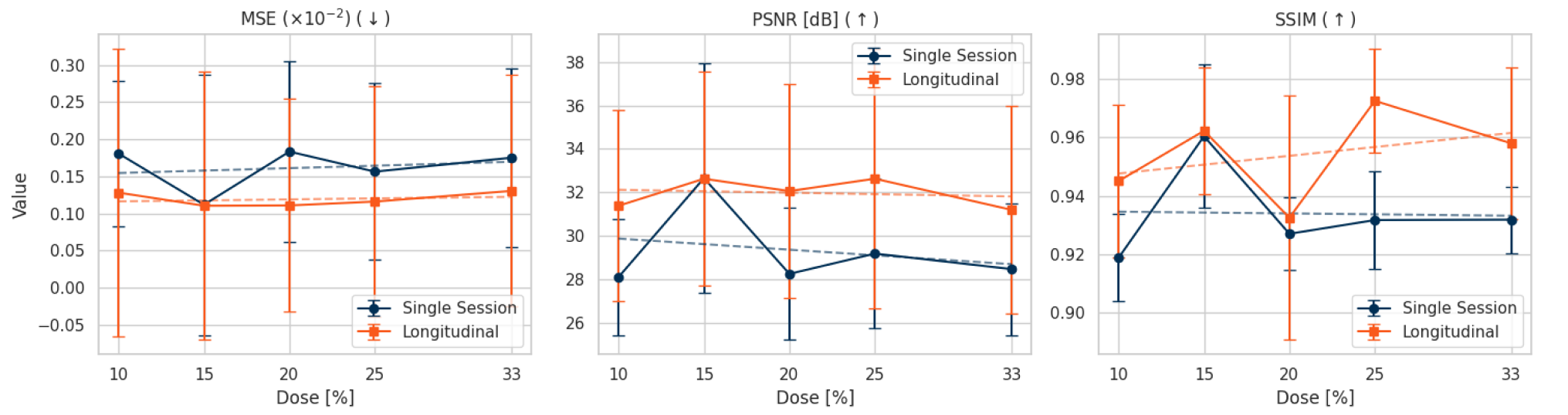}
    \caption{Effect of Simulated Dose Levels on the Virtual Contrast Enhancement Metrics. Dotted lines represent linear regression curves fitted across dose levels for each model. Error bars denote standard deviation.}
    \label{fig_dose}
\end{figure}

Figure~\ref{fig_dose} shows the virtual contrast enhancement performance for both the single session and longitudinal approaches at different dose levels (10\%, 15\%, 20\%, 25\% and 33\%).
Interestingly, at a dose of 15\%, both models converge in three different reconstruction metrics.
For other dose levels, however, the longitudinal model consistently outperforms the single session model in all metrics.
The regression slopes are not statistically significant (p > 0.05), indicating stable performance across doses.

\section{Discussions}

In this study, a longitudinal data-driven model was proposed with the objective of improving the performance of virtual contrast enhancement MRI models for patients undergoing multiple follow-up examinations.
The performance of the presented model was quantitatively compared to that of a single session model that had been trained exclusively on second-session images.
The longitudinal model demonstrated significant improvements in image fidelity and structural similarity across a range of reconstruction metrics.
The findings emphasise the efficacy of incorporating longitudinal data for virtual contrast enhancement tasks and the potential for reducing contrast dose.

In this work, only T1-weighted sequences were utilised to train the model.
Although prior non-longitudinal studies have investigated the incorporation of supplementary MRI modalities, including T2-FLAIR and ADC~\cite{ammari2022can,bone2021contrast}, these studies did not demonstrate a significant improvement in performance with the incorporation of these sequences~\cite{bone2021contrast}.
Nevertheless, the integration and evaluation of such additional modalities in the context of longitudinal analyses may warrant further investigation.

As the public dataset did not include real low-dose images, synthetic low-dose inputs were generated based on the method described in~\cite{pinetz2023faithful}.
The utilization of synthetic data offers a certain degree of flexibility, allowing for controlled variation of dose levels to simulate a wide range of conditions and to augment the dataset. This approach would not be feasible within the constraints of clinical protocols in medical research.
However, it should be noted that these synthesized images do not fully replicate the complexity and variability of actual low-dose acquisitions, and several factors could limit the reliability of the results obtained using them.
In order to ensure clinical relevance and robustness, further validation should be performed with real-dose images, particularly in the context of clinical deployment.

While the proposed model demonstrates promising robustness across a wide range of time intervals (16–312 days), significant anatomical and perfusion changes, particularly between pre- and postoperative scans, can restrict the effectiveness of longitudinal data.
Such variability may affect the model’s ability to leverage prior data effectively. 
Although the framework was designed to handle moderate inter-session changes, pre- and post-operative paired data were not experimented with in this study.
Its performance under such conditions remains to be investigated.

It is important to note that the test set utilised in this study comprises only 17 patients, all obtained from a single publicly available dataset.
Despite its limited size, it includes patients with treated lesions that demonstrate a range of longitudinal responses, such as growth, shrinkage, and stability.
This heterogeneity introduces valuable clinical variability, thereby  enabling the model to be evaluated in a more realistic setting.
Nevertheless, the restricted number and source of test cases may limit the generalizability of the findings.
It is therefore recommended that future research include larger, multi-institutional cohorts with the aim of validating the proposed approach further across a range of lesion types and treatment responses.

\section{Conclusion}
The present study proposes a proof-of-concept longitudinal deep learning approach for virtually enhancing contrast of low-dose T1-weighted MRI acquisitions to generate full-dose post-contrast images. This approach utilises prior high-dose scans to enhance image fidelity and has been demonstrated to significantly improve performance in comparison to a conventional non-longitudinal single session model across a range of quantitative reconstruction metrics.
Furthermore, the approach demonstrates robustness across varying synthetised contrast levels, thereby confirming its adaptability to different clinical scenarios.
These findings emphasize the potential of incorporating longitudinal information into contrast dose reduction strategies and demonstrate the feasibility of safer, lower-dose MRI protocols without compromising diagnostic quality.
Clinically speaking, patients with chronic diseases who require regular and frequent gadolinium injections represent a key population for whom implementing protocols with alternating full and low-dose acquisitions may offer a favorable risk-benefit profile.

\subsubsection{\ackname}
This work was granted access to the HPC resources of IDRIS under the allocation 2024-A0150314655 made by GENCI. 
This work was supported by ANRT (CIFRE 2023/1206).

\subsubsection{\discintname}
P.F., A.B., N.D., P.R. are employees of Guerbet.

\bibliographystyle{splncs04}

\end{document}